\documentstyle[12pt]{article}
\begin{document}
\thispagestyle{empty}
\begin{center}{AN AMENDED FORMULA FOR THE DECAY OF RADIOACTIVE MATERIAL FOR
COSMIC TIMES}\end{center}
\vspace{1cm}
\begin{center}{Moshe Carmeli}\end{center} 
\begin{center}{Department of Physics, Ben Gurion University, Beer Sheva 84105, 
Israel}\end{center}
\begin{center}{(E-mail: carmelim@bgumail.bgu.ac.il)}\end{center}
\begin{center}{and}\end{center}
\begin{center}{Shimon Malin}\end{center}
\begin{center}{Department of Physics and Astronomy, Colgate University, 
Hamilton,\newline New York 13346, USA}\end{center}
\begin{center}{(E-mail: SMALIN@CENTER.COLGATE.EDU)}\end{center}
\vspace{1cm}
\begin{center}{ABSTRACT}\end{center}
An amended formula for the decay of radioactive material 
is presented. It is a 
modification of the standard exponential formula. The new formula applies for
long cosmic times that are comparable to the Hubble time. It reduces to the 
standard formula for short times. It is shown that the material decays faster
than expected. The application of the new formula to direct measurements of
the age of the Universe and its implications is briefly discussed. \newpage
\begin{center}{1. INTRODUCTION}\end{center}
In this paper we present an amended formula for the decay of radioactive 
material for cosmic times, when the times of the decay are of the order of 
magnitude of the Hubble time. It is reduced to the
standard formula for short times. To this end we proceed as follows.

We assume, as usual, that the probability of disintegration during any interval of 
{\it cosmic} time $dt'$ is a constant,
$$\frac{dN}{dt'}=-\frac{1}{T'}N,\eqno(1a)$$ 
in analogy to the standard formula 
$$\frac{dN}{dt}=-\frac{1}{T}N,\eqno(1b)$$
where $T'$ is a constant to be determined in terms of the half-lifetime $T$ 
of the decaying material.

It has been shown that the addition of two cosmic times $t_1$ backward with 
respect to us (now), and $t_2$ backward with respect to $t_1$, is not just 
$t_1+t_2$. 
Rather, it is given by [1-5]
$$t_{1+2}=\frac{t_1+t_2}{1+t_1t_2/\tau^2}.\eqno(2)$$
In Eq.(2) $\tau$ is the Hubble 
time in the limit of zero gravity, and thus it is a universal constant. 
Equation (2) is the universal formula for the addition of cosmic times, and 
reduces to the standard formula of times $t_{1+2}=t_1+t_2$ for short
times with respect to $\tau$.\vspace{2mm}\newline 
\begin{center}{2. DERIVATION OF THE FORMULA}\end{center}
Let us substitute in the formula for the addition of cosmic times, Eq.(2),
$t_1=-t$ and $t_2=-dt$. Then
$$t_{1+2}=-\frac{t+dt}{1+tdt/\tau^2}\approx -\left(t+dt\right)\left(1-
\frac{tdt}{\tau^2}\right)\approx -\left[t+dt\left(1-\frac{t^2}{\tau^2}
\right)\right].\eqno(3)$$
Accordingly
$$-\left(t+dt\right)\rightarrow -\left[t+dt\left(1-\frac{t^2}{\tau^2}\right)
\right],\eqno(4)$$
or 
$$-dt\rightarrow -dt\left(1-\frac{t^2}{\tau^2}\right).\eqno(5)$$

So far the times denoted backward times. Since radioactivity deals with 
forward times, we use now the standard notation of times, and Eq.(5) will be
written as
$$dt\rightarrow dt'=dt\left(1-\frac{t^2}{\tau^2}\right).\eqno(6)$$
Equation (1a) will thus have the form 
$$\left(1-\frac{t^2}{\tau^2}\right)^{-1}\frac{dN}{dt}=-\frac{1}{T'}N.\eqno(7)$$

The solution of Eq.(7) is then given by 
$$N=N_0\exp\left[-\frac{t}{T'}\left(1-\frac{t^2}{3\tau^2}\right)\right],
\eqno(8)$$ 
in analogy to the solution of the standard equation (1b), 
$$N=N_0\exp\left(-\frac{t}{T}\right).\eqno(9)$$
\begin{center}{3. DETERMINING $T'$ IN TERMS OF HALF-LIFE TIME $T$}\end{center}
From the solution (9) we have 
$$N\left(T\right)=N_0/e,\eqno(10)$$
where $T$ is the half-life time of the material, as expected. From Eq.(8), we
obtain 
$$N\left(T'\right)=N_0\exp\left[-\left(1-\frac{T'^2}{3\tau^2}\right)\right]
=\left(N_0/e\right)\exp\frac{T'^2}{3\tau^2}.\eqno(11)$$
Using Eq.(10), we now have 
$$N\left(T'\right)=N\left(T\right)\exp\frac{T'^2}{3\tau^2}.\eqno(12)$$
Under the assumption that $T'\neq 0$, we thus have 
$$N\left(T'\right)>N\left(T\right).\eqno(13)$$

In order to determine $T'$ in terms of $T$, we procede as follows. We 
substitute in Eq.(8) $t=T$, and using Eq.(10),we obtain
$$N\left(T\right)=N_0\exp\left[-\frac{T}{T'}\left(1-\frac{T^2}{3\tau^2}\right)
\right]=N_0/e.\eqno(14)$$
As a result we have 
$$\frac{T}{T'}\left(1-\frac{T^2}{3\tau^2}\right)=1,\eqno(15)$$
or
$$T'=T\left(1-\frac{T^2}{3\tau^2}\right),\eqno(16)$$
and thus 
$$T'<T.\eqno(17)$$
Using Eq.(16) in Eq.(8) we therefore obtain
$$N\left(t\right)=N_0\exp\left[-\frac{t\left(1-t^2/3\tau^2\right)}
{T\left(1-T^2/3\tau^2\right)}\right].\eqno(18)$$ 
Accordingly we have
$$N\left(t\right)=N_0\exp\left[\frac{-t\alpha\left(t\right)}{T}\right],
\eqno(19)$$
where 
$$\alpha\left(t\right)=\frac{1-t^2/3\tau^2}{1-T^2/3\tau^2}\geq 1;\hspace{1cm} 
(t\leq T).\eqno(20)$$
Also we have 
$$N_0\exp\left[\frac{-t\alpha\left(t\right)}{T}\right]\leq 
N_0\exp\left(-\frac{t}{T}\right).\eqno(21)$$

Consequently, 
Eq.(19) provides a large deviation 
from Eq.(9) when $T$ is comparable to $\tau$ and we measure radioactivity
over astronomical times. For example, Thorium is a radioactive element with a
half-life of 14.1 billion years, as compared to the estimated 13 billion years
age of the Universe. Such measurements/observations can 
be carried out, and the detected deviations can be drawn by a graph (see Fig. 1).  
In principle, it follows from Eq.(19) that $N(t)$ for a given $t$ is less than
that obtained through the traditional formula; i.e., the material decays 
faster than expected.
\vspace{2mm}\newline
\begin{center}{4. DISCUSSION}\end{center}
Accurate measurements for the decay of radioactive materials from the Earth 
and from stars in 
our galaxy, could provide crucial information about the age
of the Universe. It is well known that two of the most straightforward methods 
of calculating the age of the Universe -- through redshift measurements, and 
through stellar evolution -- yield incompatible results. Recent measurements
of the distances of faraway galaxies through the use of the Hubble Space
Telescope indicate an age much less than the ages of the oldest stars that we 
calculate through the stellar evolution theory [6-16]. 

At present there is no conclusion to this contraversy; a cosmological constant
would probably verify the situation, but it is possible that the discrepancy 
will 
disappear with more accurate measurements of the age of the Universe using 
both methods. The discussion given in this paper clearly goes in the right 
direction in solving this important impasse.
\newpage
\begin{center}{REFERENCES}\end{center}
1. M. Carmeli, {\it Found. Phys.} {\bf 25}, 1029 (1995).\newline
2. M. Carmeli, {\it Found. Phys.} {\bf 26}, 413 (1996).\newline
3. M. Carmeli, {\it Inter. J. Theor. Phys.} {\bf 36}, 757 (1997).\newline
4. M. Carmeli, {\it Cosmological Special Relativity: The Large-Scale Structure
of Space, Time and Velocity}, World Scientific (1997), p. 24.\newline
5. M. Carmeli, Aspects of cosmological relativity, in: {\it Proceedings of the
Fourth Alexander Friedmann International Seminar on Gravitation and Cosmology}, 
held in St. Peterburg, June 17-25, 1998,
Russian Academy of Sciences Press, in print.\newline
6. W.L. Freedman {\it et al.}, {\it Nature} {\bf 371}, 757 (1994).\newline
7. W.L. Freedman, HST highlights: The extragalactic distance scale, in: {\it
Seventeenth Texas Symposium on Astrophysics and Cosmology}, H. B\"{o}hringer, 
G.E. Morfill and J.E. Tr\"{u}mper, Editors, Annals of the New York Academy of
Sciences, New York, Vol. 759 (1995), p. 192.\newline 
8. A. Renzini, in: {\it Observational Tests of Inflation}, T. Shanks 
{\it et al.}, Editors, Kluwer, Boston (1991).\newline
9. A.R. Sandage, {\it Astr. J.} {\bf 106}, 719 (1993).\newline
10. X. Shi, D.N. Schramm, D.S.P. Dearborn and J.W. Truran, {\it Comments on
Astrophysics} {\bf 17}, 343 (1995).\newline
11. D.E. Winget {\it et al.}, {\it Astrophys. J.} {\bf 315}, L77 (1987).\newline
12. E. Pitts and R.J. Taylor, {\it Mon. Not. R. Astr. Soc.} {\bf 255}, 557 
(1992).\newline
13. W. Fowler, in: {\it 14th Texas Symposium on Relativistic Astrophysics},
E.J. Fenyores, Editor, New York Academy of Sciences, New York (1989), p. 68.\newline 
14. D.D. Clayton, in: {\it 14th Texas Symposium on Relativistic Astrophysics}, 
E.J. Fenyores, Editor, New York Academy of Sciences, New York (1989), p. 79.\newline
15. S.M. Carroll, W.J. Press and E.L. Turner, {\it A. Rev. Astr. Astrophys.}
{\bf 30}, 499 (1992).\newline
16. M.J. Pierce {\it et al.}, {\it Nature} {\bf 371}, 385 (1994).
\newpage
\begin{center}{CAPTIONS}\end{center}
Fig. 1: Two curves describing the standard exponential decay $N/N_0=\exp\left(
-t/T\right)$ and the amended cosmic decay $N/N_0=\exp\left[-t\alpha\left(t
\right)/T\right]$. For a measured $N/N_0$ the two curves occur at two
different times $t_1$ and $t_2$, with $t_2>t_1$, where $t_1$ and $t_2$ 
correspond to the amended and the standard decay formulas. Accordingly, 
cosmic times of decaying materials on Earth and stars are actually shorter 
than has been believed so far.
\end{document}